\documentclass[aps,prd,showpacs,amssymb,twocolumn]{revtex4}

\usepackage{graphicx}% Include figure files
\usepackage{dcolumn}% Align table columns on decimal point
\usepackage{bm}% bold math

\begin{document}
%\LARGE
%\preprint{gr-qc/}
%\draft
\title{Birefringence in nonlinear anisotropic dielectric media}
\author{V. A. De Lorenci}
 \email{delorenci@unifei.edu.br}
\author{R. Klippert}
 \email{klippert@unifei.edu.br}
\author{D. H. Teodoro}
 \email{dhteodoro@unifei.edu.br}
\affiliation{Instituto de Ci\^encias Exatas,
Universidade Federal de Itajub\'a, 
Av.\ BPS 1303 Pinheirinho, 37500-903 Itajub\'a, MG, Brazil} 

\date{September 22, 2004}
%\twocolumns[
%\hsize\textwidth\columnwidth\hsize\csname@twocolumnfalse\endcsname 

\begin{abstract}
Light propagation is investigated in the context of local anisotropic 
nonlinear dielectric media at rest with the dielectric coefficients 
$\varepsilon^\mu{}_\nu = \varepsilon^\mu{}_\nu (\vec{E},\vec{B})$
and constant $\mu$, in the limit of geometrical optics. Birefringence
was examined and the general conditions for its occurrence were presented. 
A toy model is exhibited, in which uniaxial birefringent media
with nonlinear dielectric properties could be driven by external fields in such
way that birefringence may be artificially controlled. The effective
geometry interpretation is also addressed.
\end{abstract}

%]
\maketitle

%\begin{multicols}{2}

%
\section{Introduction}
\label{I}
%\hspace{0.5cm} 
%
In the regime of intense electromagnetic fields or inside some material media
the equations governing electrodynamic phenomena are nonlinear. In the first 
case, the theory is built from a nonlinear Lagrangian \cite{Plebanski} which 
is a function of the two Lorentz invariants of the electromagnetic 
field. In the second case, 
Maxwell equations must be supplemented with constitutive 
relations between external applied fields and the induced excitations. 
In general, such relations are nonlinear (although linear constitutive relations 
are also of interest \cite{Rubilar}), and they depend on the physical 
properties of each considered medium under the action of external fields.  
In both cases the field equations will be presented in a nonlinear form
and, as a consequence, several non usual effects (in the context of Maxwell theory) 
are predicted. 

Of particular interest is the phenomenon of 
birefringence: the light velocity dependence on the polarization mode of the 
propagating wave.
In the case of nonlinear electrodynamics, birefringence occurs when the 
electric and/or magnetic fields exceed its critical value %($m_e^2c^2/e\hbar$) 
predicted by quantum
electrodynamics \cite{Schwinger}. The analysis of light propagation in such
regime shows that there is a probability of photon splitting 
under a strong external electromagnetic field \cite{Birula}.
More recent investigations on light propagation in the context of nonlinear
Lagrangian for electrodynamics can be found in Refs.\ 
\cite{Dittrich,DeLorenci,Novello,Refere1} and references therein. Particularly, 
in Ref.\ \cite{Refere1} the Fresnel analysis of wave propagation in nonlinear
electrodynamics was performed. For a class of local nonlinear Lagrangian nondispersive
models, birefringence phenomena were studied and the wave propagation 
in a moving isotropic nonlinear medium was also examined.

Natural uniaxial birefringence is a well known effect \cite{Born}
and it takes place for some materials (mostly crystals). Artificial
birefringence is also possible to occur \cite{Landau,Souza,Klippert,prl2}, 
as an induced effect in material media: when an external field is applied in a medium with 
nonlinear dielectrics properties, an artificial optical axis may appear. 
Birefringence is also used as a technique for investigating other 
properties of some systems: see for instance Ref. \cite{prl3}, where the birefringence effect 
is used as a tool for astrophysical studies.

In this work, birefringence is investigated in the context of homogeneous
dielectric media at rest with the dielectric coefficients 
$\varepsilon^\mu{}_\nu = \varepsilon^\mu{}_\nu (\vec{E},\vec{B})$ 
and constant $\mu$ in the limit of geometrical optics. The analysis is restricted
to local electrodynamical models, and dispersive effects were neglected 
by considering only monochromatic waves, 
thus avoiding ambiguities with the velocity of the wave. A mechanism 
by which birefringence can be controlled by means of an external electric 
field is  proposed. In particular, it is shown that naturally uniaxial
media presenting nonlinear dielectric properties can be operated 
by external fields in such way that birefringence could be artificially 
turned off.

In the following section, the generalized eigen-vector equation associated with the light propagation 
in general anisotropic media is presented. In Sec.\ \ref{III} the solution for such 
equation is presented for the particular case where the impermeability $\mu^\mu{}_\nu$
is assumed to be a constant, and  birefringence conditions are 
examined. Section \ref{IV} particularizes to the case $\varepsilon^\alpha{}_\beta(E)$, 
and the principal refractive indices are explicitly presented. 
In Sec.\ \ref{V} a particular model is 
exhibited in which a naturally uniaxial media presenting nonlinear dielectric
properties, under the action of an external electric field, behaves in such way that 
birefringence could to be artificially controlled.
 
A covariant formalism is used throughout this work. Spacetime is assumed to be Minkowskian,
and a Cartesian coordinate system is used, such that the metric is 
$\eta_{\mu\nu} = {\rm diag}(+1,-1,-1,-1)$. 
Units are chosen such that $c=1$, except mentioned otherwise.  
A geodetic observer $V^\mu=\delta^\mu_0$
is supposed to describe all quantities. Particularly the electric field is 
represented by $E^\mu = -F^{\mu\nu}V_\nu = (0,\,\vec{E})$ whose modulus 
is $E=(-E^\alpha E_\alpha)^{1/2}$, and similarly for the magnetic field 
$B^\mu=(0,\,\vec{B})$.

%=========================================================================================

\section{Wave equation}
\label{II}

The electrodynamics in a medium at rest is completely determined by the Maxwell
equations
\begin{eqnarray}
V^{\mu}D^{\alpha}{}_{,\mu}+\eta^{\alpha\beta\gamma\delta}
V_{\gamma}H_{\delta,\beta}&=&0
\label{9}
\\
V^{\mu}B^{\alpha}{}_{,\mu}-\eta^{\alpha\beta\gamma\delta}
V_{\gamma}E_{\delta,\beta}&=&0,
\label{10}
\end{eqnarray}
together with the constitutive relations %\cite{Hely}
\begin{eqnarray}
D^{\alpha} &=&
\varepsilon^{\alpha}{}_{\beta}E^{\beta}
\label{1}
\\
H^{\alpha} &=& \mu^{\alpha}{}_{\beta}B^{\beta},
\label{2}
\end{eqnarray}
where the coefficients $\varepsilon^{\alpha}{}_{\beta}=\varepsilon^{\alpha}{}_{\beta}
(\vec{E},\,\vec{B})$ and  $\mu^{\alpha}{}_{\beta}=\mu^{\alpha}{}_{\beta}
(\vec{E},\,\vec{B})$ represent the dielectric tensors (called 
permittivity and impermeability tensors, respectively) which encompass 
all information about the dielectric properties of the medium. 
They are usually assumed as being frequency-dependent, although they may more generally 
depend on the external fields themselves. The coordinate derivatives of 
Eqs. (\ref{1}) and (\ref{2}) are
\begin{eqnarray}
D^\alpha{}_{,\mu}\! &=&\! \varepsilon^{\alpha}{}_{\beta}E^\beta{}_{,\mu}
\!+\! \frac{\partial \varepsilon^\alpha{}_{\beta}}{\partial
E^\tau}E^{\beta}E^\tau{}_{,\mu} \!+\! \frac{\partial
\varepsilon^{\alpha}{}_{\beta}}{\partial B^\tau}E^{\beta}B^\tau{}_{,\mu}
\label{3}
\\
H^\alpha{}_{,\mu}\! &=&\! \mu^{\alpha}{}_{\beta}B^\beta{}_{,\mu}
\!+\!\frac{\partial \mu^{\alpha}{}_{\beta}}{\partial E^\tau}
B^{\beta}E^\tau{}_{,\mu}
\!+\!\frac{\partial \mu^{\alpha}{}_{\beta}}{\partial
B^\tau}B^{\beta}B^\tau{}_{,\mu}.
\label{4}
\end{eqnarray}

In order to determine the propagation of the electromagnetic waves, 
we will consider the eikonal approximation of electrodynamics,
making use of the method of field discontinuities \cite{Plebanski,Hadamard}. 
With the notation introduced in Ref. \cite{Souza}, we set   
%\begin{equation}
$\left[E^\mu{}_{,\nu}\right]_{\Sigma} = e^{\mu}K_{\nu}$ and
%\;\;\;\;
$\left[B^\mu{}_{,\nu}\right]_{\Sigma} = b^{\mu}K_{\nu},$
%\label{35}
%\end{equation}
where $e^{\mu}$ and $b^{\mu}$ represent the polarization of 
the wave fields on the surface of discontinuity 
$\Sigma$ and $K_{\lambda}=\partial \Sigma / \partial x^\lambda$ 
is the wave vector normal to $\Sigma$.
By applying these boundary conditions to the Eqs. (\ref{9}) and 
(\ref{10}), and taking into account Eqs. (\ref{3}) and (\ref{4}), we obtain 
$b^\mu = \eta^{\mu\nu\lambda\sigma}K_\nu V_\lambda e_\sigma / \omega$, and
\begin{eqnarray}
&&\left(\omega C^{\alpha}{}_\tau
+\frac{\partial\mu_{\delta\lambda}}{\partial E^\tau}
\eta^{\alpha\beta\gamma\delta} B^{\lambda} V_\gamma K_\beta +
\right. \nonumber \\ 
&&\left.
+\frac{1}{\omega} 
H_\delta{}^{\chi} \eta^{\alpha\beta\gamma\delta} \eta_{\chi\lambda\nu\tau}
V_\gamma K^\lambda V^\nu K_\beta \right)e^\tau = 0,
\label{16}
\end{eqnarray}
where we introduced the new quantities
\begin{eqnarray}
C^{\alpha}{}_\tau &\doteq&\varepsilon^{\alpha}{}_{\tau} +
\frac{\partial \varepsilon^{\alpha}{}_{\beta}}{\partial E^{\tau}}E^\beta + 
\frac{1}{\omega}\frac{\partial \varepsilon^{\alpha}{}_{\beta}}{\partial B^{\rho}}
\eta^{\rho\lambda\gamma}{}_{\tau}E^{\beta}K_\lambda V_\gamma,
\label{14}
\\
H^{\alpha}{}_\tau &\doteq& \mu^{\alpha}{}_{\tau} + \frac{\partial 
\mu^{\alpha}{}_{\beta}}{\partial B^{\tau}}B^{\beta},
\label{15}
\end{eqnarray}
with $\omega \doteq K^\alpha V_\alpha$ the frequency of the
electromagnetic wave.

In order to present Eq. (\ref{16}) in a 3-dimensional representation we introduce
the projector on the 3-space $h^\alpha_\mu = \delta^\alpha_\mu - V^\alpha V_\mu$,
and define 
\begin{equation}
q^\alpha = h^\alpha_\mu K^\mu = K^\alpha - \omega V^\alpha.
\label{17}
\end{equation}
Straightforward simplifications lead Eq. (\ref{16}) to the form
\begin{equation}
Z^\alpha{}_\tau e^\tau=0,
\label{21b}
\end{equation}
where 
\begin{eqnarray}
Z^\alpha{}_\tau = C^{\alpha}{}_\tau
+\frac{q}{\omega}\frac{\partial\mu_{\delta\lambda}}{\partial E^\tau}
\eta^{\alpha\beta\gamma\delta} B^{\lambda} V_\gamma \hat{q}{}_\beta 
%\\
%\nonumber
%+\frac{1}{\omega} \left( q^2 I^{[\omega}{}_\tau H_\omega{}^{\alpha]}
%+H_\tau{}^{\beta}q_\beta q^\alpha - H_\chi{}^{\delta}q^\chi q_\delta h^{\alpha}{}_\tau \right)
+\frac{q^2}{\omega^2}I^{[\alpha}_\beta I^{\lambda]}_\tau H_\lambda{}^\beta %,
\label{21}
\end{eqnarray}
in which $a^{[\mu\nu]} = a^{\mu\nu} - a^{\nu\mu}$ for an arbitrary tensor 
$a^{\mu\nu}$. Further, we have defined the projector 
\begin{equation}
I^\alpha_\tau=h^\alpha_\tau +\hat{q}{}^\alpha \hat{q}{}_\tau,
\label{22}
\end{equation}
on the 2-space orthogonal to the propagation direction
$\hat{q}^\mu\doteq q^\mu/q$, where $q^2=-h_{\mu\nu}q^\mu q^\nu$.
 
The general solution for the wave propagation can be derived from 
the eigen-vector problem stated by Eq.\ (\ref{21b}), and is given by
the generalized Fresnel equation 
\begin{equation}
\det|Z^\alpha{}_\beta| = 0.
\label{32}
\end{equation}
For several physical configurations \cite{Novello} the dispersion relations coming from 
Eq.\ (\ref{32}) can be written in the suggestive form
$g^{\mu\nu}_{\pm}K^\pm_\mu K^\pm_\nu =0$, as shown in Sec.\ \ref{III}.

%=========================================================================================

\section{Birefringence conditions}
\label{III}

Let us assume the more specific case where the medium behaves in such way that
$\mu^\alpha{}_\beta=\mu^{-1}h^\alpha_\beta$ with $\mu$ a constant. 
In this case, Eq. (\ref{21}) reduces to
\begin{equation}
Z^\alpha{}_\tau=C^{\alpha}{}_\tau - \frac{1}{\mu v_\varphi^2} I^\alpha{}_\tau,
\label{25}
\end{equation}
where $v_\varphi = \omega/q$ is the phase velocity of the wave.  
Equation (\ref{32}) can be written as 
$(Z^\mu{}_\mu)^3 -3 Z^\mu{}_\mu Z^\alpha{}_\beta Z^\beta{}_\alpha
+2 Z^\alpha{}_\beta Z^\beta{}_\gamma Z^\gamma{}_\alpha = 0$, 
which yields a fourth order equation of the form
\begin{equation}
\alpha v_\varphi^4 - \beta v_\varphi^2 - \gamma = 0,
\label{effective}
\end{equation}
where we have defined 
\begin{eqnarray}
\!\!\!\!\alpha \!&\doteq&\! \frac{1}{6} \left[ (C^\mu{}_\mu)^3 \!-
3 C^\mu{}_\mu C^\alpha{}_\beta C^\beta{}_\alpha
\!+2 C^\alpha{}_\beta C^\beta{}_\gamma C^\gamma{}_\alpha \right]\!,
\label{44}
\\
\!\!\!\!\beta \!&\doteq&\!  \mu^{-1} \left( C^\lambda{}_\alpha C^{\alpha\nu}
-C^\alpha{}_\alpha C^{\lambda\nu} \right)\hat{q}_\lambda \hat{q}_\nu,
\label{46}
\\
\!\!\!\!\gamma \!&\doteq&\! \mu^{-2} C^{\lambda\nu} \hat{q}_\lambda \hat{q}_\nu .
\label{47}
\end{eqnarray}
The solutions of Eq. (\ref{effective}) are
\begin{equation}
v^\pm_\varphi = \sqrt{\frac{\beta}{2\alpha}\left(1 \pm \sqrt{1 
+ \frac{4\alpha \gamma}{\beta^2}}\right)}.
\label{49}
\end{equation}
Equation (\ref{49}) expresses the fact that, in general, the velocity 
of the electromagnetic waves inside a material medium may get two possible values 
$v_\varphi^+, v_\varphi^-$  which are associated
to the two possible polarizations modes \cite{Klippert}. Thus, 
birefringence occurs if the following inequalities hold for a given direction
$\hat{q}{}^\lambda$:
\begin{eqnarray}
-\frac{1}{4} < \frac{\alpha\gamma}{\beta^2} &<& 0,
\label{64}
\\
\alpha\beta &>& 0.
\label{63}
\end{eqnarray}
%with $(v^\pm_\varphi)^2 > 0$.
(For the particular case $\varepsilon^\alpha{}_\beta=\epsilon h^\alpha_\beta$,
with $\epsilon$ a constant, one obtains $\alpha=\epsilon^3$, $\beta=2\epsilon^2/\mu$ and 
$\gamma=-\epsilon/\mu^2$, and the two phase velocities reduce both to the same well-known
value $v_\varphi ^2 = 1/\epsilon\mu$.)

%====================================================================================================

\subsection{Effective geometry}

The dispersion relation Eq.\ (\ref{effective}) can be conveniently 
expressed in the form $g^{\mu\nu}_{\pm}K^\pm_\mu K^\pm_\nu =0$ (corresponding to the 
factorization of the general fourth-order Fresnel tensor \cite{Referee2}), where 
\begin{equation}
g^{\alpha\beta}_{\pm} = \mu\alpha V^\alpha V^\beta + \frac{1}{2}\left[ C^\nu{}_\nu 
- \frac{1}{\mu (v^\pm_\varphi)^2}\right]C^{(\alpha\beta)} - \frac{1}{2}C^{(\alpha}{}_{\nu}C^{\nu\beta)} 
\label{geometry}
\end{equation}
%is the effective geometry associated to the wave propagation.  
The symmetric tensors $g^{\mu\nu}_{\pm}$ represent the effective geometries 
(also known as optical metrics) associated with the wave propagation, 
and the symbol $\pm$ indicates that, in general, 
there will be two possible distinct metrics, one for each polarization mode. 
The integral curves of the vectors $K^\pm_{\mu}$ are geodesics in the 
corresponding effective geometry $g^{\mu\nu}_{\pm}$ \cite{Klippert}. 
For the particular case of the linear theory in vacuum, 
both $g^{\mu\nu}_+$ and $g^{\mu\nu}_-$ reduce to $\eta^{\mu\nu}$ as expected. 
For the case of naturally isotropic nonlinear media at rest 
the results presented in Refs.\ \cite{Souza,Refere1,Klippert} are recovered. 

The effective geometry interpretation for electromagnetic waves in material media
\cite{Souza,Klippert,Brevik} (or even in the context of nonlinear electrodynamics in vacuum
\cite{Dittrich,Novello,Gutierrez}) has been proposed as a tool for testing kinematic 
aspects of general relativity in laboratory \cite{analog}. 
In this context, Eq. (\ref{geometry}) provides a natural support 
for analogue anisotropic models.

%====================================================================================================

\section{Non-magnetic anisotropic nonlinear media}
\label{IV}

Let us consider a naturally uniaxial medium that
reacts nonlinearly when subjected to an external electric field as
$\varepsilon^\alpha{}_\beta = {\rm diag}[0,\varepsilon_{\parallel} 
(E),\varepsilon_{\perp} (E),\varepsilon_{\perp} (E)]$. 
In this case, $\varepsilon^\alpha{}_\beta=\varepsilon^\alpha{}_\beta(E)$ 
and Eq.\ (\ref{14}) takes the simpler form 
\begin{equation}
C^{\alpha}{}_\beta =\varepsilon^{\alpha}{}_{\beta} -
\frac{\partial \varepsilon^{\alpha}{}_{\tau}}{\partial E}
\frac{E^\tau E_\beta}{E}.
\label{89}
\end{equation}
By setting $\vec{E}$ in the $x$-direction (optical axis), we obtain
$C^\alpha{}_\beta={\rm diag}(0,\varepsilon_{\parallel}+ E \varepsilon'_{\parallel},
\varepsilon_{\perp},\varepsilon_{\perp})$, 
where $\varepsilon'_{\parallel} = d\varepsilon_{\parallel} / dE$. 
Now, from the above results, we obtain
\begin{eqnarray}
\alpha\gamma = -\frac{\varepsilon_\perp^2 C^1{}_1}{\mu^2}\left[
\varepsilon_\perp(1-\hat{q}_1{}^2) + C^1{}_1 \hat{q}_1{}^2\right],
\label{106-a}
\\
4 \alpha \gamma + \beta^2=\left[\frac{\varepsilon_\perp}
{\mu}(C^1{}_1 -\varepsilon_\perp )(1-\hat{q}_1{}^2)\right]^2,
\label{106}
\\
\alpha\beta=\frac{\varepsilon_\perp{}^3}{\mu}C^1{}_1\left[C^1{}_1(1+\hat{q}_1{}^2) 
+\varepsilon_\perp(1-\hat{q}_1{}^2)\right]>0,
\label{110}
\end{eqnarray} 
which show that conditions (\ref{64}) and (\ref{63}) are fulfilled in this case, provided that 
$C^1{}_1 \neq \varepsilon_\perp$ and $\hat{q}_1{}^2<1$, 
since both $C^1{}_1$ and $\varepsilon_\perp$ are positive quantities.  

Equation (\ref{106}) shows that birefringence can be annulled if 
$C^1{}_1=\varepsilon_\perp$. In other words, birefringence phenomena could 
be turned off by imposing that
\begin{equation}
\varepsilon_\parallel+E \varepsilon'_\parallel = \varepsilon_\perp,
\label{112}
\end{equation}
which is a {\em no-birefringence condition}. Note that if a model for dielectric 
permittivity is set, the no-birefringence condition
will mean a condition on the value of the external electric field.

\subsection{Ordinary and extraordinary rays}

By considering the previous model,  Eq. (\ref{49}) yields
\begin{eqnarray}
(v^\pm_{\varphi})^2=&&\frac{1}{2\mu\varepsilon_\perp C^1{}_1}\left[C^1{}_1(1+\hat{q}_1{}^2) +
\varepsilon_\perp (1-\hat{q}_1{}^2) \right. +
\nonumber \\
&&\left. \pm (C^1{}_1-\varepsilon_\perp)(1-\hat{q}_1{}^2)\right].
\label{124}
\end{eqnarray}
Thus, there will be two phase velocities, one of which is isotropic (ordinary ray) 
\begin{equation}
(v_\varphi^+)^2=\frac{1}{\mu\varepsilon_\perp},
\label{125}
\end{equation}
and the other one depends on the direction of propagation (extraordinary ray) 
\begin{equation}
(v_\varphi^-)^2=\frac{1}{\mu\varepsilon_\perp C^1{}_1}\left[\varepsilon_\perp 
(1-\hat{q}_1{}^2)+C^1{}_1\hat{q}_1{}^2\right].
\label{126}
\end{equation}
The two velocities coincide when either the propagation
occurs along the direction of the electric field, 
or when the no-birefringence condition Eq. (\ref{112}) holds.
 
The two extreme values of $v_\varphi^-{}$
occur in the direction of the electric field ($\theta = 0$), and perpendicularly to it
($\theta = \pi/2$). 
For these two directions, we obtain
\begin{eqnarray}
v_ \parallel {}^2 &\doteq& \left. (v_\varphi^-)^2 \right|_{\theta = 0} 
=\frac{1}{\mu\varepsilon_\perp} = (v_\varphi^+)^2,
\label{127}
\\
v_ \perp {}^2 &\doteq& \left. (v_\varphi^-)^2 
\right|_{\theta = {\pi}/{2}} =\frac{1}{\mu C^1{}_1}.
\label{128}
\end{eqnarray}
Associated to these directions we define the principal effective refractive indexes 
$n_o$ and $n_e$ by 
\begin{eqnarray}
n_o^2 &=& \frac{1}{v_ \parallel {}^2}=\mu \varepsilon_ \perp,
\label{130}
\\
n_e^2 &=& \frac{1}{v_ \perp {}^2}=\mu (\varepsilon_ \parallel
+ E\varepsilon'_ \parallel).
\label{129}
\end{eqnarray}
The effective refractive index of the extraordinary ray along
an arbitrary direction $\hat{q}{}^\lambda$ is determined by
$n_q=1/v_\varphi^-$, with $v_\varphi^-$ given by Eq.\ (\ref{126}).

%=========================================================================================

\section{Controlled birefringence}
\label{V}

Let us examine  some special cases where
birefringence could be controlled by adjusting the magnitude of an external electric
field.

%\subsection{$\varepsilon_\perp=\epsilon_\perp-3pE^2$} 
\subsection{Non-birefringent nonlinear media}

We consider the model where $\varepsilon_\perp=\epsilon_\perp
-3pE^2$, with $\epsilon_\perp$ and $p$ constants.
Thus, from Eq. (\ref{112}) we obtain
$\varepsilon_\parallel+E \varepsilon'_\parallel=\epsilon_\perp-3pE^2$. 
%\label{117}
%whose general solution is
%\begin{equation}
%\varepsilon_\parallel=\epsilon_\perp+\frac{C_0}{E}-pE^2.
%\label{118}
%\end{equation}
%By setting $C_0=0$ (note that  $C_0\neq 0$ would represent a non physical situation
%in the limit $E \rightarrow 0$), we obtain
Regularity on the solution of this equation at $E=0$ yields
\begin{eqnarray}
\varepsilon_\parallel=\epsilon_\perp-pE^2.
\label{119}
\end{eqnarray}
It means that for every value of $E$, if the medium reacts like Eq. (\ref{119}), 
there won't be birefringence phenomena.

%\subsection{$\varepsilon_\perp=\epsilon_\perp-3pE^2$ and $\varepsilon_\parallel=\epsilon_\parallel-sE^2$}
\subsection{Anisotropic media with artificially controlled birefringence}

Now, let us examine the model for which the medium reacts as
\begin{eqnarray}
\varepsilon_\perp&=&\epsilon_\perp-3pE^2,
\label{120}
\\
\varepsilon_\parallel&=&\epsilon_\parallel-sE^2,
\label{121}
\end{eqnarray}
with $\epsilon_\perp$, $p$, $\epsilon_\parallel$ and $s$ constants.
For the case $p=s$, all values of $E$ will correspond to either 
a no-birefringent configuration (if $\epsilon_\perp = \epsilon_\parallel$), 
as described by Eq. (\ref{119}), or else 
a birefringent configuration (if $\epsilon_\perp\neq\epsilon_\parallel$). 
On the other hand, for $p\neq s$, Eq. (\ref{112}) yields  
$\epsilon_\parallel-3sE^2=\epsilon_\perp-3pE^2$, 
and solving it for the electric field one obtains
\begin{equation}
E_c{}^2=\frac{\epsilon_\parallel-\epsilon_\perp}{3(s-p)}.
\label{122}
\end{equation}
Thus, from Eq. (\ref{122}) we conclude that there will be a particular 
value $E_c$ of the electric field for which the no-birefringence condition will 
be satisfied. Any other value of $E\neq E_c$ will result in birefringence.
Of course these results are only of interest in the case $E_c$ 
lies within the limit of applicability of Maxwell theory.
Further, the quadratic terms in $E$ in Eqs. (\ref{120}) and (\ref{121}) should
remain small when compared with the constant ones. Equation (\ref{122}) is 
meaningful for positive media ($\epsilon_\perp < \epsilon_\parallel$) if
$p<0$ and $s>0$; and similarly for negative media ($\epsilon_\perp > \epsilon_\parallel$)
if $p>0$ and $s<0$.
%
%
%\begin{widetext}
\begin{center}
\begin{figure}[thp]
\leavevmode
\centering
%\hspace*{-1.1cm}
\includegraphics[scale=0.8]{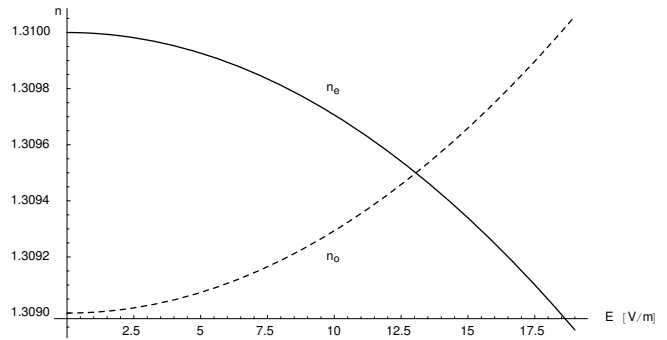}
\caption{Controlled birefringence: the particular value $E_c$ of the electric field
for which birefringence is turned off occurs in the intersection of
the curves representing the effective refractive indices for the ordinary and extraordinary
rays. The naturally uniaxial positive crystal becomes negative when the electric field
exceeds $E_c$. Note that, for $E=0$, natural birefringence remains. 
We used the following illustrative values (in MKS units) 
for the parameters appearing in Eqs. (\ref{120}) and (\ref{121}): 
$\epsilon_\parallel = 1.5195 \times 10^{-11}F/m$, 
$\epsilon_\perp = 1.5172 \times 10^{-11}F/m$,
$p = -2.2648 \times 10^{-17}Fm/V^2$ 
and $s = 2.2700 \times 10^{-17}Fm/V^2$.
\label{graph-index}
}
\end{figure}
\end{center}
%\end{widetext}
%
For the case where $\epsilon_\parallel \neq \epsilon_\perp$ birefringence will 
occur provided that $E \neq E_c$. For the particular case where $E=0$ 
we have natural birefringence, as it occurs in some crystals.  
See Fig. \ref{graph-index} for a numerical analysis of
a toy model where natural birefringence is artificially controlled by the
influence of an external electric field.

By considering a naturally isotropic medium where $\epsilon_\parallel=\epsilon_\perp$, 
the model set by Eqs.\ (\ref{120}) and (\ref{121}) 
describes artificial birefringence \cite{Souza} 
as it appears in the Kerr electro-optic effect. For this case we obtain
$n_o- n_e  \approx \lambda K E^2$, where we have defined
%\begin{equation}
%\lambda K \doteq c \sqrt{\mu\epsilon_ \perp}\frac{3(p-s)}{2\epsilon_ \perp},
%\label{134}
%\end{equation}
$\lambda K \doteq 3\sqrt{\mu\epsilon_ \perp}(s-p)/2\epsilon_ \perp$,
with $\lambda$ the wave length of the electromagnetic waves and $K$ 
the  Kerr's constant (its value depends on the dielectric properties of the medium).
The effective geometry for this particular case has already been presented
in the literature \cite{Souza,Refere1,Klippert}.

%====================================================================================================

\section{conclusion}
\label{VI}

A covariant tensorial formalism is here provided in order to discuss 
the propagation of monochromatic electromagnetic waves inside naturally anisotropic 
material media with nonlinear dielectric properties, in the limit of geometrical optics.  
The eigen-vector problem for general media was presented.  
For the case of constant $\mu$, the conditions 
for birefringence phenomena to appear  were examined.  
For the case of non-magnetic media, a particular model was shown for 
which the magnitude of the birefringence could be controlled, and even turned off, 
with the use of an external electric field. 
The model generalizes the standard description of the Kerr electro-optical effect 
to the case of a naturally anisotropic material media.  
All the obtained results can be straightforwardly rewritten to the case 
$\varepsilon^\alpha{}_\beta=\epsilon h^\alpha_\beta$ with $\epsilon=const$ and 
$\mu^\alpha{}_\beta=\mu^\alpha{}_\beta(\vec{B})$, thus generalizing the Cotton-Mouton effect 
(magnetic analogue to the Kerr effect) to the case of anisotropic media. 

The results can also be used in the context of analogue models for general relativity.  
It is here presented the optical metric for nondispersive anisotropic media, 
which extends previous investigations. 
Isotropic moving media were already discussed \cite{Refere1}, 
and their results agree with ours in the case of isotropic media at rest. 
Anisotropic moving media are still to be considered.

\acknowledgments
This work was partially supported by the Brazilian research agencies CNPq and FAPEMIG. 
DHT thanks the support from the PIBIC-UNIFEI program.

%\end{multicols}
\end{document}